# Generative AI in Financial Institution: A Global Survey of Opportunities, Threats, and Regulation

Bikash Saha, Nanda Rani, Sandeep Kumar Shukla
Department of Computer Science & Engineering,
Indian Institute of Technology Kanpur
{bikash,nandarani,sandeeps}@cse.iitk.ac.in

**Abstract**

Generative Artificial Intelligence (GenAI) is rapidly reshaping the global financial landscape, offering unprecedented opportunities to enhance customer engagement, automate complex workflows, and extract actionable insights from vast financial data. This survey provides an overview of GenAI adoption across the financial ecosystem, examining how banks, insurers, asset managers, and fintech startups worldwide are integrating large language models and other generative tools into their operations. From AI-powered virtual assistants and personalized financial advisory to fraud detection and compliance automation, GenAI is driving innovation across functions. However, this transformation comes with significant cybersecurity and ethical risks. We discuss emerging threats such as AI-generated phishing, deepfake-enabled fraud, and adversarial attacks on AI systems, as well as concerns around bias, opacity, and data misuse. The evolving global regulatory landscape is explored in depth, including initiatives by major financial regulators and international efforts to develop risk-based AI governance. Finally, we propose best practices for secure and responsible adoption – including explainability techniques, adversarial testing, auditability, and human oversight. Drawing from academic literature, industry case studies, and policy frameworks, this chapter offers a perspective on how the financial sector can harness GenAI's transformative potential while navigating the complex risks it introduces.

**Keywords**. Generative AI, Generative AI in Finance, Financial Services, Financial Sector AI Regulation, Adversarial AI Attacks, Financial Fraud and AI, Ethical AI Governance.

1. **Introduction**

Generative AI, exemplified by large language models such as GPT-4 [1] and domain-specific models like BloombergGPT [2], has emerged in recent years as a transformative technology at the forefront of innovation. Unlike traditional AI systems focused on prediction or classification, GenAI can create new content (text, code, images, etc.) based on learned patterns, enabling natural interactions and complex problem-solving. Financial institutions are eagerly exploring GenAI to reimagine services, driven by its success in other domains and the competitive pressure to innovate [7] [17]. The potential upside is enormous: the McKinsey Global Institute projects that generative AI could add $200–340 billion annually to the banking sector by boosting productivity and revenues [3]. A 2024 PwC India survey similarly found that 90% of Indian financial firms are focusing on AI/GenAI for innovation, especially to enhance customer experience and decision-making [4].

At the same time, financial services operate in a highly regulated, risk-sensitive environment, so unbridled use of AI carries dangers. If not properly managed, GenAI could misjudge human nuances or generate faulty outputs, leading to customer harm or operational errors [3]. The technology's ability to convincingly mimic human communication also raises security red flags – for example, AI-generated deepfakes and fake news have already caused market disruptions [5]. Moreover, financial AI systems must comply with strict requirements for fairness (e.g. in lending decisions) and privacy (handling of sensitive data). These concerns have prompted a flurry of activity by regulators and banks to establish guardrails even as adoption accelerates.

This chapter provides a global survey of GenAI in financial ecosystems, balancing enthusiasm with caution. We first document how banks and fintech organizations worldwide are adopting generative AI across diverse use cases and examine the benefits and value propositions—ranging from enhanced customer service to improved compliance—that are driving this trend. Next, we discuss emerging cybersecurity threats enabled by GenAI and new attack vectors targeting AI systems along with frameworks like MITRE ATLAS that map these risks. We also address ethical and governance issues such as bias, explainability, and data privacy which are critical in finance. The next section reviews the regulatory landscape shaping AI usage in finance, including initiatives by Reserve Bank of India (RBI), Securities and Exchange Board of India (SEBI), United States Securities and Exchange Commission (U.S. SEC), European Union (EU), and Monetary Authority of Singapore (MAS). We conclude with recommendations for secure AI adoption, advocating risk management frameworks, robust testing (including red teaming), transparency tools, and human oversight to ensure that generative AI's deployment remains safe, fair, and effective. An overview of the sections is shown in Figure 1. Through this comprehensive overview, readers will gain insight into how to harness GenAI's transformative power in financial ecosystems while responsibly managing its risks.

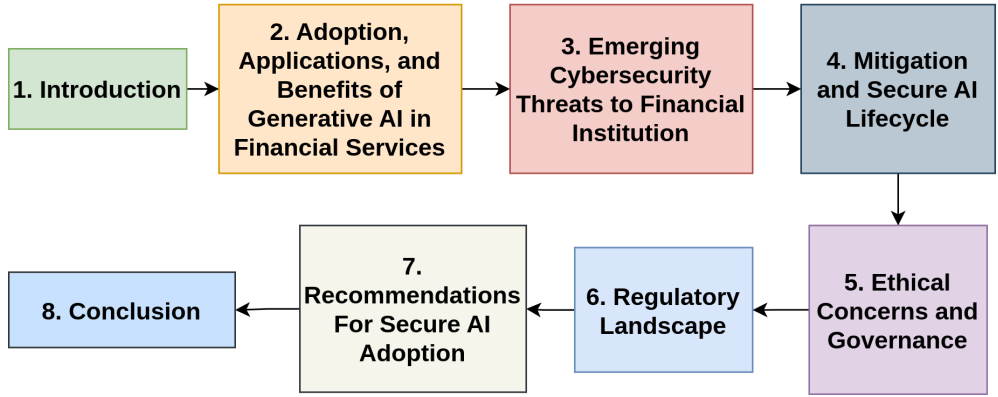

**Fig 1:** Section Overview

## 2. ADOPTION AND APPLICATIONS OF GENERATIVE AI IN FINANCIAL SERVICES

The financial services industry is witnessing a profound shift with the integration of Generative Artificial Intelligence (GenAI) technologies. Across banking, investment management, regulatory compliance, and internal operations, GenAI is redefining how institutions engage with customers, manage risks, optimize operations, and drive strategic decision-making [6]. Based on the classification in Figure 2, this section outlines key GenAI application domains, their use cases, and the underlying technical approaches.

### 2.1. Customer-Facing Functions

One of the earliest and most prominent areas of GenAI adoption is in customer-facing functions [7]. Banks and financial institutions are increasingly deploying GenAI-driven chatbots and virtual assistants to handle a broad range of customer inquiries with human-like fluency [8]. Fine-tuned large language models (LLMs), often supplemented by multilingual embeddings, have enabled institutions such as the State Bank of India (SBI) and Axis Bank to pilot AI chatbots capable of serving diverse customer bases across multiple regional languages [66] [67]. These systems improve customer satisfaction and also contribute to operational efficiencies by reducing the workload on human service agents [9].

Further advancements in customer engagement involve the generation of personalized financial recommendations. Leveraging Retrieval-Augmented Generation (RAG) techniques, financial institutions like HDFC Bank use GenAI models [68] to generate customized investment insights, wealth management advice, and product suggestions tailored to individual client profiles [10] [11]. On the marketing front, GenAI facilitates dynamic campaign content creation through prompt chaining and text-to-text generation methods, enabling more targeted customer outreach and adaptive marketing strategies [12].

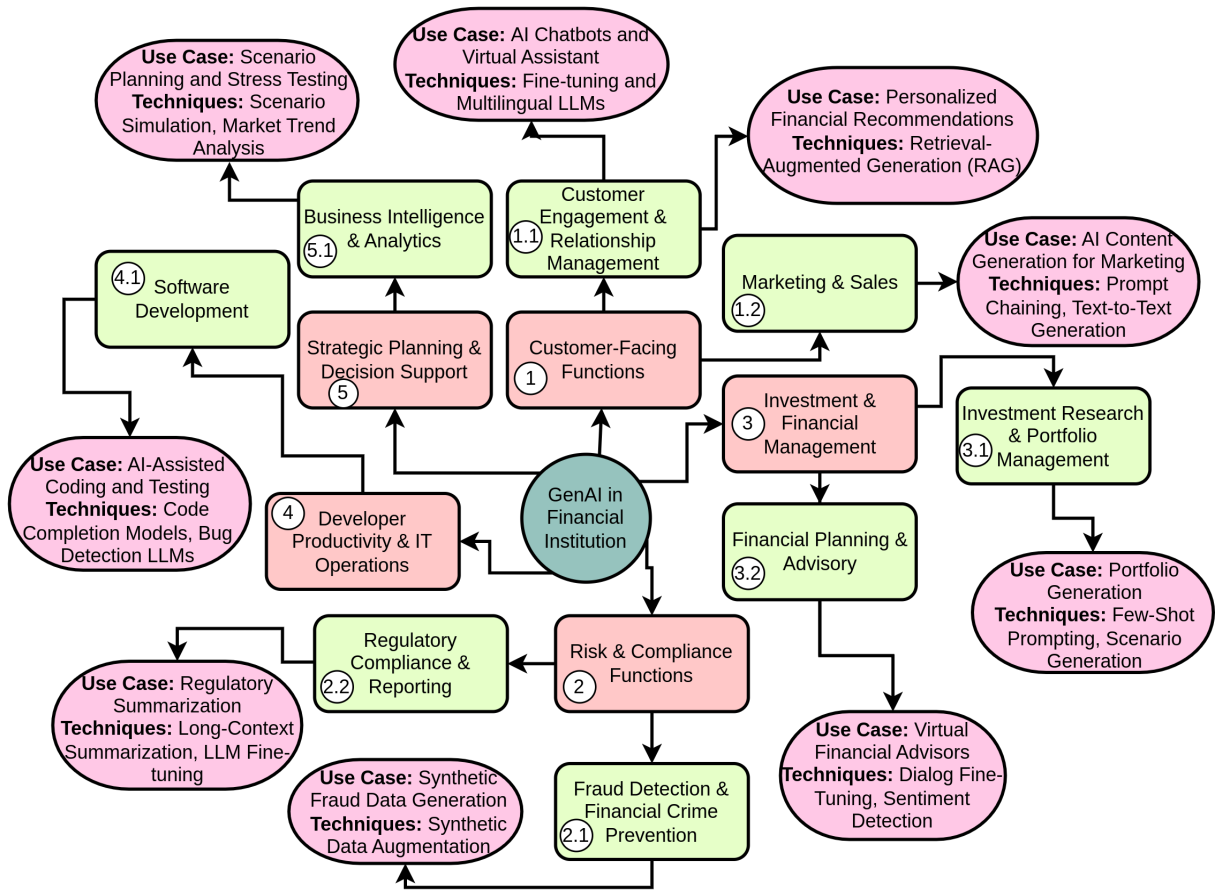

**Fig 2:** Classification of Application of GenAI in Financial institution

## *2.2.  Risk and Compliance Functions*

Risk management and regulatory compliance represent critical domains where GenAI is delivering measurable value. Financial crime detection is being enhanced through the generation of synthetic fraud scenarios. The studies have employed synthetic data augmentation to create extensive fraud training datasets, enabling fraud detection systems to anticipate novel fraudulent behaviors that traditional systems might miss [13] [14] [15] [16].

Startups like AdvaRisk, leverages its GenAI-driven data intelligence platform to transform collateral management, offering comprehensive solutions for financial institutions [70].

In the compliance arena, leading institutions like Citigroup discuss operationalizing GenAI systems to automate regulatory document summarization and reporting [69]. Utilizing long-context summarization techniques and fine-tuned LLMs, these systems parse lengthy regulatory texts, extract actionable obligations, and generate compliance narratives, significantly reducing manual workload and improving regulatory agility [17]. The capacity of GenAI to handle dense legal and compliance documentation aligns with findings from recent studies indicating that LLM-based summarizers outperform traditional NLP models in comprehension and extraction tasks within finance [7] [18].

### 2.3. *Investment and Financial Management*

In investment management, GenAI models are being employed to generate customized portfolios and provide real-time investment insights. JPMorgan's IndexGPT is an operational example wherein users can create thematic portfolios through natural language queries, enabled by few-shot prompting and scenario generation techniques [19].

Financial planning and advisory services are also evolving through the deployment of virtual financial advisors. These systems, piloted in innovation labs of various banks, use dialogue fine-tuning and sentiment detection to simulate human-like advisory conversations, offering personalized guidance while maintaining scalability [20]. Recent research has highlighted that GenAI-driven advisory platforms significantly enhance customer trust and advisory satisfaction compared to earlier rule-based systems [21] [22].

### 2.4. *Developer Productivity and IT Operations*

Generative AI has found substantial application within internal development and IT operations. Financial institutions are increasingly integrating AI coding assistants to support software engineers [23]. Organizations such as Goldman Sachs have reported productivity gains by embedding GenAI models capable of autocompleting code, detecting bugs, and generating test cases, particularly for complex financial software systems [24] [25]. Code completion models and bug detection LLMs have been instrumental in accelerating software development cycles, improving code reliability, and reducing human error. This trend reflects broader findings in the financial technology sector where GenAI-enabled development support is correlated with a 20–30% reduction in time-to-market for new applications [26].

## 2.5. *Strategic Planning and Decision Support*

Strategic planning and business intelligence functions are leveraging GenAI's capacity for unstructured data mining and scenario simulation. Financial institutions utilize scenario planning models that generate plausible economic and market outcomes, facilitating stress testing of investment portfolios and operational resilience planning [27]. Techniques such as market trend analysis and simulated scenario generation allow decision-makers to anticipate risks and opportunities more comprehensively than traditional forecasting methods. Recent empirical studies confirm that financial firms using GenAI-assisted strategic tools exhibit superior adaptability to market shocks and regulatory changes compared to firms relying solely on conventional data analytics [28].

Across all domains of financial services, several cross-cutting patterns characterize the adoption of Generative AI. Institutions typically initiate GenAI integration through pilot deployments focused on improving internal productivity, such as implementing employee-facing chatbots or AI-assisted coding tools, before gradually scaling to customer-facing or production-grade applications [29]. Rather than replacing human decision-making, GenAI is predominantly employed in an augmented intelligence model, where it functions as a cognitive co-pilot, enhancing human expertise and accelerating complex analytical and operational tasks [29]. To align GenAI with the specialized needs of financial services, techniques such as Retrieval-Augmented Generation (RAG), synthetic data augmentation, long-context document summarization, and dialogue fine-tuning are frequently adapted and optimized for domain-specific applications. Furthermore, the importance of regulatory compliance, transparency, and ethical AI governance is increasingly emphasized, with institutions actively establishing AI governance frameworks to mitigate risks and ensure responsible innovation [30]. Generative AI thus emerges not merely as a technological enhancement but as a strategic enabler across the financial services ecosystem. Its applications span customer relationship management, fraud detection, regulatory compliance, investment advisory, software development, and strategic planning, contributing to operational resilience and competitive advantage [31]. As financial institutions deepen their integration of GenAI, success will be determined by their ability to tailor techniques to specific operational contexts while maintaining a steadfast commitment to governance and trustworthiness. Future research should focus on longitudinal evaluations of GenAI's impact on operational robustness, regulatory adherence, financial inclusion, and sustained customer trust within the evolving digital finance landscape.

## 3. EMERGING CYBERSECURITY THREATS TO FINANCIAL INSTITUTION

While generative AI unlocks value, it also introduces new cybersecurity threats and amplifies existing ones. These threats fall into two broad categories: **(a) threats enabled by GenAI** – where malicious actors use generative AI to enhance their attacks – and **(b) threats targeting AI systems** deployed by organizations – exploiting vulnerabilities in the AI models or supply chain. These treats are further classified as shown in Figure 3.

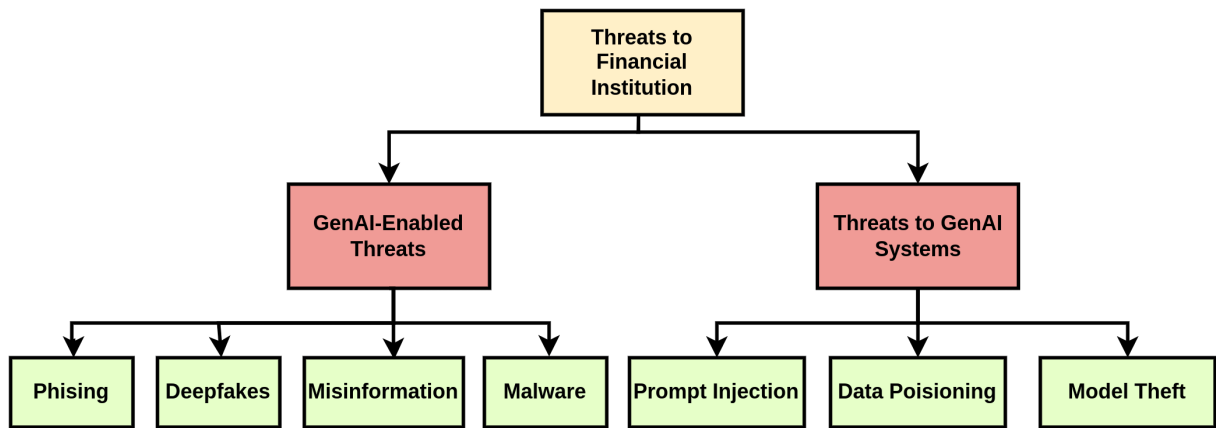

**Fig 3:** Threat Categorization to Financial Institutions

In the financial sector, which deals with sensitive data and funds, the stakes of these threats are especially high. We discuss major emerging threats and frameworks for understanding and mitigating them:

### 3.1. GenAI-Enabled Threats

The rise of Generative AI (GenAI) has significantly amplified attackers' capabilities across **phishing, social engineering, deepfakes, misinformation, and malware development** [31] [32] [33]. Financial institutions are witnessing a surge in AI-generated phishing attacks, where emails and messages, crafted with near-flawless grammar and mimicking specific individuals or organizations, bypass traditional detection cues. A 2024 industry report noted a 118% rise in AI-driven phishing and deepfake activities, with phishing emerging as the top global fraud tactic [34]. Highly personalized phishing emails, often using social media data, have increased the success rate of business email compromise (BEC) attempts, where AI-generated communications impersonate CEOs or vendors, sometimes supported by deepfake voice calls [71].

Deepfake technology poses an equally severe threat. Attackers now create synthetic audio or video to impersonate executives or clients, authorizing fraudulent financial transactions. In one study, 26% of executives targeted by deepfake scams reported that the attacker's goal was to trigger unauthorized transfers [35]. AI-generated misinformation can also manipulate financial markets, as demonstrated by the brief stock market dip following a fake AI-generated image of an explosion near the Pentagon in May 2023 [5]. Financial institutions increasingly face the risk of AI-fueled disinformation campaigns, including forged documents and manipulated public sentiment, prompting agencies like FinCEN to issue alerts on the use of deepfakes in financial fraud [36].

Moreover, GenAI has lowered barriers to malware development. Previously requiring specialized skills, malware creation is now aided by malicious AI models like "WormGPT" [37] and "FraudGPT" [38], which are sold on underground forums without ethical safeguards. These black-hat AI tools enable criminals to generate phishing sites, malicious code, and customized exploits at scale, creating polymorphic malware that can evade traditional security solutions. FraudGPT, for instance, was demonstrated generating a functional phishing website impersonating Bank of America in seconds. Security researchers have also found that open-source AI models, if fine-tuned on malware data, can produce ransomware and keyloggers [72].

Overall, GenAI is intensifying the cybersecurity threat landscape, making phishing attacks more convincing, social engineering more potent, disinformation more disruptive, and malware more accessible. Financial institutions must urgently upgrade defenses, adopting AI-driven threat detection, implementing multi-factor authentication beyond voice verification, and enhancing employee and customer training against AI-enabled fraud. Despite growing awareness, the threat remains substantial, with 90% of companies reporting cyber-fraud targeting in 2024, driven heavily by AI-enhanced techniques [34]. The cybersecurity arms race against GenAI threats has clearly begun, demanding an evolution in both technological defenses and human vigilance.

*3.2.      Threats Targeting GenAI Systems*

As financial firms increasingly deploy generative AI systems, they must defend not only against AI-enabled attacks but also against attacks targeting their own AI models. Generative AI introduces unique vulnerabilities that adversaries can exploit, including prompt injection, data extraction, and supply chain manipulation. Prompt injection attacks occur when an attacker embeds hidden instructions into user input or external data sources, causing the AI to override safety guardrails and perform unintended actions. Researchers have demonstrated how indirect prompt injections, such as hidden text on websites, could hijack AI assistants like ChatGPT's plugins to extract personal data or alter system behaviors [39] [40]. For financial institutions, the risk extends to AI-powered transactional systems, where a successful prompt injection could lead to unauthorized trades or data breaches. Closely related is the threat of model inversion or data extraction, where attackers systematically query AI models trained on sensitive internal data to extract confidential client information, trading strategies, or proprietary insights [41]. Financial AI models fine-tuned on internal datasets are particularly vulnerable to such leaks, even if the data is revealed through subtle inference rather than direct disclosure.

Supply chain attacks, particularly data poisoning and malicious model insertion, pose an additional serious risk. In data poisoning, attackers corrupt the training or fine-tuning datasets to introduce vulnerabilities or biases. A poisoned fraud detection model, for instance, might learn to ignore certain fraudulent patterns, making targeted attacks easier to execute. Demonstrations like "PoisonGPT," [42] where researchers uploaded a subtly sabotaged LLM to a public repository, highlight the feasibility of such attacks [38] [41]. Financial institutions relying on third-party models or open-source datasets face heightened risks if model integrity is not rigorously verified through checksums, provenance tracking, or model signing. In parallel, model theft or replication, where attackers reconstruct proprietary AI systems through extensive querying, represents a growing threat, particularly for valuable trading or credit risk models. Adversarial inputs—specially crafted queries designed to fool AI decision-making—also pose risks, such as tricking a loan approval system into approving high-risk applicants.

The consequences of these attacks are profound, potentially resulting in financial loss, reputational damage, privacy violations, and even systemic market disruptions if critical

AI-driven trading or risk management systems are compromised. Recognizing the emerging threat landscape, frameworks like MITRE ATLAS have been developed to systematically catalog AI-specific adversarial tactics, such as LLM Prompt Injection, Data Poisoning, and Model Extraction [43]. ATLAS maps technical attack vectors and also provides case studies, such as the ChatGPT plugin leak and PoisonGPT, demonstrating real-world vulnerabilities. In response, the cybersecurity community advocates best practices like AI Bills of Materials (AI BOMs), model provenance tracking, and mandatory model signing to ensure AI integrity throughout its lifecycle [38]. As generative AI becomes integral to financial operations, securing AI systems against these sophisticated adversarial threats will be critical to maintaining institutional trust, operational resilience, and market stability.

4. **MITIGATION AND SECURE AI LIFECYCLE**

In response to the emerging threats against AI systems, financial institutions are adopting a structured secure AI development lifecycle, embedding security principles and risk controls systematically across every phase of AI development, deployment, and operation [44] [45] [46]. The demonstrate a sample lifecycle in Figure 4.

This lifecycle approach is designed to anticipate, resist, detect, and respond to adversarial threats targeting generative AI models, ensuring that AI systems can function safely in high-risk financial environments.

   a) **Secure Data Collection and Model Training:** The first phase begins with secure data collection and model training. Financial institutions place emphasis on sourcing curated, trusted training data, applying version control and cryptographic verification to prevent data poisoning attacks. Adversarial training techniques are employed during model development, where models are deliberately exposed to malicious prompt patterns and evasion attempts, training them to resist exploitation [47]. This ensures that the AI learns not only functional tasks but also adversarial resilience from the outset.

   b) **Rigorous Validation and Red Teaming:** The second phase focuses on rigorous validation and red teaming. Before deployment, AI systems undergo extensive adversarial testing through simulated attacks conducted by internal teams or external experts. Red-teaming exercises, as advocated by frameworks like MITRE ATLAS, help reveal vulnerabilities such as prompt injection susceptibility, model inversion risks, or behavioral drift under adversarial inputs [48]. Financial institutions increasingly mandate vendors to demonstrate adversarial resilience evidence before integrating third-party models into operational workflows.

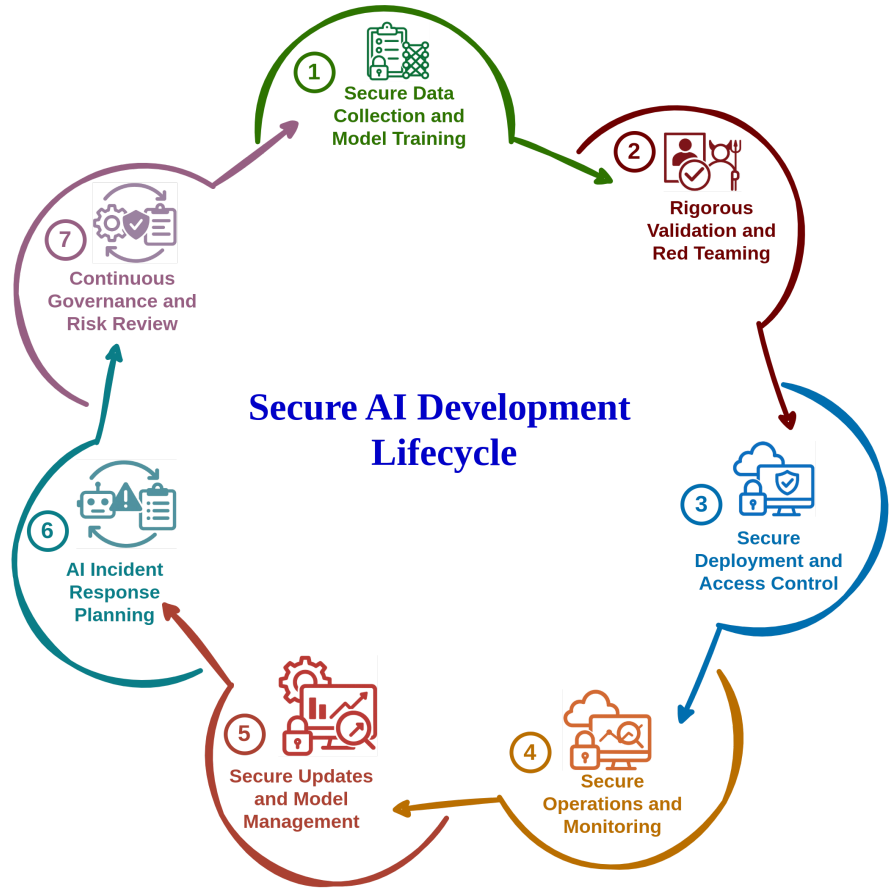

**Fig 4:** Secure AI Development Lifecycle

c) **Secure Deployment and Access Control:** Once validated, AI systems move to the deployment and access control phase. In this phase, robust user authentication, input sanitization, and strict authorization protocols are implemented [49]. AI-driven systems such as GPT-based financial assistants are gated through access controls and usage monitoring, while human-in-the-loop oversight is required for sensitive actions like financial transactions or data disclosures. Real-time logging and anomaly detection systems are deployed to monitor AI behavior continuously, allowing organizations to detect and halt anomalous outputs indicative of compromise or manipulation.

d) **Secure Operations and Monitoring:** Operational security is further reinforced during the secure operations phase. AI system interactions are logged

      comprehensively, and automated alerting mechanisms flag abnormal behaviors, such as unexpected data exposure or anomalous decision outputs [50]. In high-risk cases, sessions can be terminated automatically, and human supervisors review flagged incidents for escalation, mirroring intrusion detection systems used in traditional cybersecurity.

  e) **Secure Updates and Model Management:** Managing model updates and retraining activities securely forms the next critical phase [51]. Financial institutions treat AI models as sensitive software artifacts, storing them encrypted and digitally signed. Before any retraining or model update is deployed to production, extensive validation is performed to detect any unexpected behaviors or vulnerabilities that could have been introduced. Additionally, when third-party models are adopted, their cryptographic hashes are verified against known standards to ensure model provenance and prevent supply chain poisoning attacks.

  f) **AI Incident Response Planning:** Recognizing that no system is invulnerable, financial firms also augment their incident response frameworks to handle AI-specific breaches [52]. Predefined response steps are established for AI incidents, including isolation of compromised AI systems, log analysis, impact assessment, and model patching or retraining as needed. Cross-functional teams, blending cybersecurity and AI engineering expertise, are created to continuously review vulnerabilities, analyze emerging threats, and coordinate incident management efforts.

  g) **Continuous Governance and Risk Review**: Finally, the lifecycle is sustained through continuous governance and risk review [53]. Institutions align their practices with frameworks such as the U.K.'s National Cyber Security Centre (NCSC) guidelines for securing machine learning pipelines and the U.S. NIST AI Risk Management Framework. Regular audits, independent assessments, and proactive risk evaluations ensure that AI deployments remain resilient as attacker capabilities evolve. Best practices like maintaining AI Bills of Materials (AI BOMs) and implementing model signing standards are increasingly incorporated to enhance transparency and trustworthiness.

By operationalizing a complete secure AI lifecycle, financial organizations aim to create layered, defense-in-depth architectures for AI, balancing the transformative benefits of GenAI with the critical need for security and resilience. Institutions that proactively secure their AI systems across this lifecycle will be best positioned to harness generative AI while minimizing exposure to novel and sophisticated threats. Conversely, those neglecting these controls may face heightened risks of financial loss, reputational harm, and systemic disruptions fueled by AI-driven vulnerabilities.

## 5. ETHICAL CONCERNS AND GOVERNANCE

The adoption of GenAI in financial services presents technical challenges along with significant ethical and governance concerns [7] [53] [54]. Financial institutions, given their direct impact on individuals' economic opportunities, must ensure that AI systems uphold fairness, transparency, privacy, and accountability. Table 1 summarizes the key ethical concerns associated with GenAI in financial services, the risks they entail, and the mitigation strategies adopted by leading institutions.

**Table 1:** Ethical Concerns, Risks, and Mitigation Strategies for GenAI in Financial Services

| Ethical Concern | Associated Risk | Mitigation Strategy |
|---|---|---|
| **Bias and Fairness** | Discriminatory outcomes in lending, credit scoring, or fraud detection | Fairness audits, disparate impact testing, adversarial debiasing, synthetic data balancing |
| **Transparency and Explainability** | Opaque AI decisions undermine trust and regulatory compliance | Use of SHAP, LIME, interpretable models, mandated explanations for customer-impacting decisions |
| **Data Privacy and Consent** | Unauthorized data usage, privacy violations, regulatory breaches | Data anonymization, pseudonymization, explicit consent, federated learning, synthetic data generation, privacy audits |
| **Accountability and Human-in-the-Loop (HITL)** | Unchecked AI outputs causing financial errors or ethical violations | Human review of critical AI outputs, staff training to avoid automation bias, accountability frameworks |
| **Ethical Use and Societal Impact** | Employment disruption, misinformation spread, unrealistic customer expectations | Staff retraining programs, AI disclosure policies, training AI to admit uncertainty, avoidance of hallucinated outputs |

These concerns are increasingly formalized into structured governance frameworks aligned with regulatory expectations and societal trust imperatives. We discuss these issues and emerging best practices for governance:

a. **Bias and Fairness:** AI models, including GenAI, often inherit societal biases embedded within training data, risking discriminatory outcomes in areas like lending, credit scoring, or fraud detection [54]. Institutions are therefore instituting fairness audits, disparate impact testing, and bias mitigation strategies such as adversarial debiasing and synthetic data balancing to ensure compliance with equal opportunity laws and to close financial inclusion gaps. Beyond compliance, financial institutions recognize that proactively improving fairness through AI offers new opportunities to expand services to historically underserved demographics.

b. **Transparency and Explainability:** The complexity of GenAI models creates challenges in interpreting and justifying their decisions, which conflicts with the regulatory and customer demands for transparency [54]. Financial firms are integrating explainability tools like SHAP and LIME to interpret AI decisions, ensuring customers and auditors can understand the rationale behind outcomes. In sensitive contexts such as loan denials or investment recommendations, institutions are embedding mechanisms that force AI systems to produce understandable explanations, thereby preserving accountability and trust.

c. **Data Privacy and Consent:** These are paramount, given the sensitive financial data handled by AI systems [54]. With regulations like the GDPR and India's DPDP Act mandating strict data protections, institutions are employing measures such as data anonymization, pseudonymization, secure data handling, and explicit consent protocols for training AI on customer data. Techniques like federated learning and synthetic data generation are gaining adoption to reduce the privacy risks associated with centralized data training. Privacy audits are becoming routine to ensure ongoing compliance and prevent unauthorized data exposures.

d. **Accountability and Human-in-the-Loop (HITL):** These mechanisms ensure that AI outputs do not operate unchecked [7]. Regulatory guidance emphasizes that ultimate responsibility for AI-driven decisions must rest with humans, leading banks to integrate HITL designs where human officers review high-impact AI outputs and periodically audit decision patterns. Institutions are also combatting automation bias through employee training, stressing the need for critical oversight of AI recommendations, particularly in high-risk areas like fraud detection and lending.

   e. **Ethical Use and Societal Impact:** Broader ethical considerations also come into play [55]. Responsible AI adoption includes mitigating employment disruptions by retraining affected staff, ensuring customers are aware when interacting with AI systems, and preventing GenAI systems from generating fabricated or misleading outputs. Financial firms are training AI systems to acknowledge uncertainty rather than hallucinating incorrect information, thereby promoting responsible information dissemination.

To institutionalize these principles, many financial organizations are establishing AI governance frameworks and ethics committees that oversee AI projects from conception through deployment. Initiatives like Singapore's Veritas Consortium and India's RBI FREE-AI committee are developing methodologies and standards to audit AI systems across Fairness, Ethics, Accountability, and Transparency (FEAT) dimensions [60] [61]. Adherence to international principles, such as those from the OECD (Organisation for Economic Co-operation and Development) [30], further anchors these frameworks globally.

In conclusion, ethical governance of GenAI in finance is not peripheral but central to sustainable adoption. Institutions embedding fairness, explainability, privacy, accountability, and continuous governance into the AI lifecycle not only mitigate legal and reputational risks but also gain a competitive advantage through stronger consumer trust and regulatory confidence. In a future shaped by AI, ethical stewardship will distinguish leaders from laggards in the financial ecosystem.

## 6. REGULATORY LANDSCAPE

The rapid rise of AI in financial services has prompted regulators across the globe to react, seeking to ensure innovation does not outpace oversight. Financial regulators are concerned with safeguarding stability, consumer protection, fairness, and market integrity in the face of AI-driven changes [56]. In this section, we outline the evolving regulatory landscape, focusing on key global efforts, including the United States, European Union, India, and Singapore, among others. The trend is towards a mix of **principles-based guidelines,** **specific rules for high-risk AI uses, and collaborative sandboxes** to test AI innovations under supervision.

### 6.1. Emerging Governance Models

#### 6.1.1. Principles-Driven, Risk-Based Frameworks

Jurisdictions like India, Singapore, and the UK are building AI regulation on foundational principles of fairness, accountability, transparency, and ethics. India's FREE-AI (Framework for Responsible and Ethical AI) committee—formed by the Reserve Bank of India (RBI)—reflects this model [60]. It aims to study AI adoption in finance and recommend guardrails, covering concerns like auditability, consumer redressal, model bias, and concentration risks.

In Singapore, the Monetary Authority of Singapore (MAS) launched Project MindForge, a multi-phase initiative to co-develop a GenAI governance framework with the financial industry [61]. MAS had earlier released the FEAT principles and continues to champion a collaborative approach, encouraging industry experimentation under responsible innovation norms. Other regions, such as Canada, Japan, and the UK, are similarly relying on existing financial regulations while issuing guidance tailored to AI.

#### 6.1.2. Codified Regulation for High-Risk Use

The European Union has taken the lead in codifying AI regulation through the forthcoming AI Act, a risk-based legislation that classifies AI systems into Unacceptable, High, Limited, and Minimal Risk categories [61]. AI systems used in credit scoring, insurance underwriting, or customer profiling are deemed high-risk, requiring rigorous data governance, transparency, human oversight, cybersecurity, and conformity assessments. The AI Act also mandates that such systems be registered in an EU-wide database, and compliance will be supervised by existing financial regulators. Financial institutions using general-purpose models like GPT-4 will also need to ensure that providers meet accountability and documentation requirements. In parallel, the EU Digital Operational Resilience Act (DORA), effective in 2025, complements the AI Act by enforcing strict ICT and third-party risk management, especially relevant for cloud-based GenAI deployments [63].

*6.1.3. Regulatory Sandboxes and Supervision Technology (SupTech)*

Around the world, regulatory sandboxes have become a preferred method to test GenAI innovations in a supervised environment. The RBI's fintech sandbox cohorts, including those for AI and ML, allow controlled experimentation and help feed insights into policy. MAS also encourages experimentation through regulated pilots and is exploring "guardrails" for GenAI in areas like customer service, model validation, and content reliability. Regulators themselves are increasingly deploying AI for supervisory purposes. For example, the US SEC uses machine learning to detect insider trading and accounting fraud. In India, the RBI is exploring AI for stress testing and anomaly detection in banking operations.

*6.1.4. Targeted Rules for Financial Use Cases*

While countries like the United States haven't enacted AI-specific laws yet, agencies such as the U.S. Securities and Exchange Commission (SEC) and Consumer Financial Protection Bureau (CFPB) have proposed and enforced rules targeting conflicts of interest, robo-advisors, and black-box discrimination [64]. The SEC's proposed "AI conflict of interest rule" mandates that financial advisors and brokers using AI must neutralize any conflict that places firm interests above clients'. Regulators emphasize that existing fiduciary duties, fairness obligations, and consumer protection laws fully apply to AI-driven decision-making. The Securities and Exchange Board of India (SEBI) has similar views. While it hasn't issued AI-specific guidelines yet, its frameworks for algorithmic trading and robo-advisory services already impose obligations on firms using AI for investment decisions. SEBI has highlighted the need for transparency, explainability, and customer suitability in AI-driven services, signaling that future AI guidelines may build upon these foundations.

*6.1.5. AI and Data Protection Convergence*

AI regulation is deeply intertwined with data protection laws. For example, India's Digital Personal Data Protection Act (2023) mandates user consent, purpose limitation, and fairness in automated decision-making—a framework highly relevant for AI in finance [65]. Similarly, the EU's GDPR and the proposed AI Act reinforce users' rights to explanations and recourse when decisions are made by AI, especially in credit-related contexts. As GenAI tools increasingly process personal financial data, global regulators are reinforcing the need for explainability, right to human review, and clear consent mechanisms to comply with privacy laws and ethical standards.

### *6.2. Toward Global Harmonization*

Global consistency remains a challenge. Cross-border financial institutions must navigate a patchwork of rules—EU's strict regime, US's sectoral guidance, and Asia's flexible, principles-based models. However, trends are emerging toward convergence:

   a. Multinational banks are pre-emptively aligning with the EU AI Act, applying its principles globally to ensure consistency and reduce compliance friction.
   b. International bodies like IOSCO (International Organization of Securities Commissions), BIS (Bank for International Settlements), and GFIN (Global Financial Innovation Network) are pushing for shared standards on AI governance, model risk, and auditability.
   c. The extraterritorial reach of regulations like the AI Act (which applies to any entity providing AI systems used in the EU) is driving a de facto globalization of AI compliance.

### *6.3. Looking Ahead*

As of 2025, no unified global AI regulation exists, but the trajectory is clear. Financial regulators are:

   a. Encouraging innovation through safe experimentation (sandboxes, pilot programs).
   b. Creating accountability mechanisms for high-impact AI applications.
   c. Clarifying obligations around fairness, transparency, and consumer rights.
   d. Collaborating internationally to avoid regulatory fragmentation.

Financial institutions globally must proactively implement governance frameworks, document AI use cases, perform model audits, and embed human oversight to stay ahead of evolving regulations. Beyond regulatory compliance, such measures enhance institutional trust, operational resilience, and strategic advantage in a GenAI-powered future.

7. **RECOMMENDATIONS FOR SECURE AI ADOPTION**

Given the opportunities of generative AI and the attendant risks outlined, financial institutions must adopt a strategic and disciplined approach to integrating GenAI into their operations [7]. This final section provides recommendations for practitioners – from banks and fintechs to insurers and asset managers – on how to securely and responsibly deploy generative AI. The recommendations combine technical measures, governance processes, and cultural practices, drawing on emerging standards and expert guidance. These recommendations are shown in Fig 5 and discussed below:

a. **Adopt an AI Risk Management Framework:** Organizations should manage AI risks with the same rigor as other enterprise risks. A good starting point is implementing frameworks like the **NIST AI Risk Management Framework (AI RMF)** [66]. The NIST AI RMF (1.0 released in 2023) offers a structured approach with functions such as **Map -> Measure -> Manage -> Govern** to identify and mitigate AI risks throughout the AI lifecycle. By mapping context and intended use, firms can understand how an AI application could go wrong; by measuring, they assess things like bias or performance drift; by managing, they put controls in place; and governance wraps around to ensure oversight and continuous improvement. An internal AI risk framework should cover ethical risks (bias, fairness), operational risks (model failure, cyber-attack), compliance risks, and reputational risks of AI.

   It's recommended to establish an AI Governance Committee or Working Group that includes stakeholders from risk, compliance, IT, business units, and data science. This body can develop internal AI policies (e.g., defining what AI use cases are permissible, setting standards for testing and validation) and review major AI projects. Additionally, integrating AI risk management into existing Model Risk Management (MRM) programs is advised. Many banks already have MRM processes per regulatory guidance; these should be updated to explicitly account for complexities of GenAI, such as dynamically learning models or the difficulty of validation. Essentially, treat AI models as high-risk models by default – requiring pre-deployment validation, approval, and periodic re-validation. Document all AI systems in an inventory along with their purpose, data sources, and known limitations.

Embrace tools and methodologies from emerging standards – for example, the ISO is working on AI quality standards, and internal audit teams can be trained to audit AI systems against these frameworks.

b. **Ensure Robust Testing, Validation, and Audit Trails:** Before deploying GenAI in any critical process, conduct thorough testing under various scenarios to evaluate performance, fairness, and robustness. This includes stress-testing the model with edge cases [67] and potential adversarial inputs (does the chatbot remain secure under weird prompts? does the credit model handle novel profiles without breaking?). Validation should not be a one-time event – implement continuous monitoring to detect model drift or unexpected outputs. For instance, if a generative model that writes customer emails starts using language that doesn't match compliance standards (perhaps due to drift in input distribution), it should be flagged.

**Audit trails** are vital: maintain detailed logs of AI model outputs and the data/prompt that led to them, especially for decision-making systems. These logs help in two ways – **transparency and accountability** (if a customer or regulator asks why something happened, we can have the record) and **debugging/forensics** if something goes wrong (like tracking a faulty output back to a flawed prompt or data issue). For example, if an AI approved a fraudulent transaction, the bank should be able to trace that decision through logs, identifying whether it was due to model error or perhaps misuse via prompt injection.

Build mechanisms to **store versions of models and data** they were trained on, so any decision can be audited against the specific model version in use at that time (this is analogous to maintaining code version control – some banks already do "model versioning"). Moreover, as part of auditability, generate **m**odel fact sheets or model cards for each GenAI system – documents describing the model's intended use, performance metrics, training data, ethical considerations, and evaluation results.

Finally, consider engaging independent audits of major AI systems – either by internal audit teams with AI expertise or external auditors/consultants – to verify compliance with policies and regulations. Regulators may soon ask for audit results of AI systems, so it's prudent to build that muscle early.

c. **Utilize Explainability and Monitoring Tools:** To tackle the black-box nature of GenAI, deploy explainability tools such as SHAP, LIME, or counterfactual generators that can help interpret model decisions. For instance, if a GenAI model is used to evaluate credit risk, use SHAP values to show the top factors that influenced each score – this can then be reviewed for plausibility and shared (in simplified form) with customers or regulators to satisfy explainability expectations. In customer service bots, design them to provide reasoning paths (e.g., "I recommended this product because you mentioned X and Y needs") which can also be logged. Additionally, implement advanced monitoring: tools that watch AI outputs for bias or policy violations. Some vendors offer AI "watchdog" systems that can sit on top of LLM outputs, scanning for toxic language or sensitive data leakage.

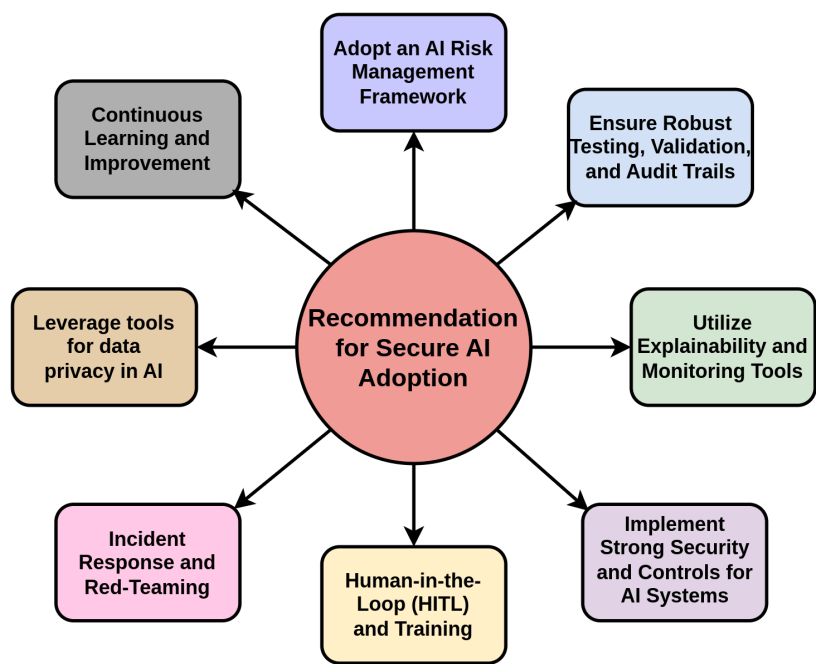

**Fig 5:** Recommendation for Secure AI Adoption

For example, before an AI-generated customer email is sent, a monitoring layer can check that it doesn't inadvertently contain any sensitive account info or inappropriate content – if it does, it flags it for human review. Similarly, drift detection tools can monitor input data characteristics over time and alert if the model is being used outside its trained scope (e.g., suddenly a credit model is seeing a demographic distribution far different from training data – possibly indicating a need to retrain or an expansion into a new market that needs careful oversight).

Bias detection tools should be run regularly: test the AI with inputs representing different genders, ethnicities, etc., and measure outcome differences. If biases are detected, have a remediation plan (adjust the model or pre/post-process the inputs/outputs to reduce bias). Explainability and monitoring are not one-time; they should be embedded in a feedback loop where model performance issues lead to model updates or policy changes. In doing so, organizations can catch issues early – for example, noticing that an AI chatbot tends to misunderstand a certain dialect and giving it special training or routing such cases to human agents.

d. **Implement Strong Security and Controls for AI Systems:** The cybersecurity controls around AI deployments should be as stringent as those for core banking systems. This includes **access control** – limit who can interact with the GenAI system, especially if it has access to sensitive data [49]. Internal AI tools should require user authentication and enforce role-based access (e.g., an AI that helps with research should not allow a retail banking clerk to query confidential investment banking research).

For AI models integrated into applications, ensure input validation and output filtering. As discussed, to mitigate prompt injections, we can sanitize inputs (strip away certain keywords or sequences) and validate outputs (e.g., if an AI is generating SQL queries as part of a process, have a rule-set to prevent dangerous commands even if the AI tried). Maintain separation of environments for development, testing, and production of AI models, with proper change management – no one should be able to tweak the model or its prompts on the fly in production without approvals.

Given the rise of threats like WormGPT, treat any externally obtained model or AI component as potentially untrustworthy: scan it for malicious content, run tests [68]. **Model signing** can ensure that only approved models (signed by the company's key) run on production infrastructure. Consider watermarking AI-generated content—such as research summaries—to enable traceability and help verify the source if misinformation arises.

On the cybersecurity team side, update threat models to include AI abuse scenarios. Train the employed SOC (Security Operations Center) analysts to recognize signs of AI-specific attacks, such as an attacker attempting many slight prompt variations to jailbreak a system (which might appear as a user rapidly inputting strange phrases – a pattern that can be detected). Also, simulate attacks (red team exercises) on AI systems to test the employed defenses and response. Essentially, **bake AI into the cyber risk program** – align it with frameworks like MITRE ATLAS and ensure the defenses cover those techniques. Regular security assessments of AI systems (penetration testing with a focus on AI endpoints, reviewing cloud configurations for AI services, etc.) are recommended.

e. **Human-in-the-Loop (HITL) and Training:** Despite automation, keep humans involved in critical decision processes. Establish clear criteria for when AI decisions must be escalated to a human. For example, a bank might say: any AI-generated loan rejection for a marginal applicant goes to a loan officer for a second look before finalizing (to prevent false negatives due to model conservatism). Or if an AI model's confidence score is below a certain threshold, it automatically flags for human review rather than acting. This can be implemented in workflows – AI provides a recommendation and a human approves or overrides. As the AI proves its accuracy over time, some thresholds might be adjusted, but **a human failsafe should always exist** for scenarios where the cost of error is high (e.g., large trading positions, significant compliance decisions).

**Training and change management** for staff is also key: employees should be trained not just in how to use new AI tools, but also in their limitations and biases. They should be encouraged to question AI output that seems off (instead of deferring to it because "the computer said so"). Many organizations now champion a culture of "human-AI teaming" – where employees know that the AI is a partner that can handle grunt work but still relies on human judgment for complex matters. This includes updating standard operating procedures: e.g., a customer service script might include instructions on when the agent should double-check the AI chatbot's answer or when to step in. A "human in the loop" approach also extends to AI development itself: include domain experts in the training process (say, having traders or loan officers provide feedback on model outputs to refine them – a form of reinforcement learning with human feedback). Continual training of AI models with human-curated data can help align the AI better with desired outcomes (like reducing hallucinations or inappropriate responses).

f. **Incident Response and Red-Teaming:** Prepare for AI-specific incidents with clear response playbooks [52]. If an AI system produces a harmful output (e.g., a privacy breach or a huge trading loss due to an AI glitch), the team should know how to triage: disable the AI if needed, inform stakeholders/regulators, patch the model or revert to an older version, etc. Incorporate AI failure scenarios into regular drills. Some firms conduct "**red team days**" where an internal team role-plays attackers trying to break an AI system's rules, while the dev team sees if they can detect and mitigate. This practice, reminiscent of cybersecurity red teaming, is gaining traction for AI assurance (OpenAI did this before releasing GPT-4, and now enterprises are adopting it). Engaging external **"AI auditors" or ethicists to red-team** the model from an ethical standpoint can also highlight issues (like finding biased outputs or ways the model could be misused) [48]. Given that regulators are concerned about AI, demonstrating that we have done such exercises can be favorable if questions arise.

g. **Leverage tools for data privacy in AI:** To adhere to privacy, consider techniques like differential privacy (adding noise to training data to prevent memorizing exact data), especially if sharing data for collaborative training [41]. Use **e**ncryption or secure enclaves if using external AI APIs – some solutions encrypt the prompt so the provider never sees raw data (although this is cutting-edge and may reduce accuracy). If possible, process sensitive data in-house: for example, use OpenAI's on-prem or dedicated instance options so that our data isn't mixed in a public pool. At minimum, utilize any settings that providers offer to not store or use our input data for their model training (OpenAI and others allow opting out, which any bank should do). Monitor compliance with data protection – ensure that any AI that requests personal

Generative AI in Financial Institution      data from a user actually needs it, and inform users (transparency notices) that AI is used and how their data is handled. Privacy and security should be part of the design, i.e., Privacy by Design and Security by Design for AI systems.

  h. **Continuous Learning and Improvement:** The field of AI and its regulation is evolving quickly. Institutions should stay updated on the latest research (e.g., new prompt injection methods and defenses), new tools (perhaps AI governance software platforms), and regulatory changes. Establish liaisons in the organization who track AI regulatory developments (like the EU AI Act, RBI and MAS guidelines) and can translate them into internal policy updates. Encourage participation in industry forums or consortiums on AI in finance, where best practices are shared. Internally, treat any incident or near-miss as a learning opportunity to harden systems. Also, collect feedback from users of AI systems (both employees and customers). If customers find an AI-driven feature confusing or unhelpful, that might indicate it needs adjustment or more human touch. Redefine KPIs (Key Performance Indicators) for AI projects to include not just efficiency but also things like customer satisfaction, error rates, and compliance metrics.

In essence, the overarching recommendation is to be deliberate and cautious in AI deployment, much like deploying a new financial product or a new core system – one wouldn't do that without extensive testing, risk assessment, and governance approvals. GenAI should be treated no differently. By putting in place strong frameworks, controls, and an organizational mindset of responsibility, financial institutions can enjoy the significant upsides of generative AI – innovation, efficiency, better insights – while keeping risks at acceptable levels. As a positive side effect, many of these practices (like better model documentation, bias testing, robust cybersecurity) will also benefit the organization's broader digital transformation and risk culture.

## 8. Conclusion

Generative AI is poised to redefine the financial landscape, offering institutions powerful tools to enhance customer service, streamline operations, and unlock data-driven insights. Although most of them are in experimental form and have started deploying pilots, there are potential use cases that we believe will be deployed. From AI-powered chatbots to investment advisory assistants, early adoption across global financial centers reals the vast potential of GenAI to revolutionize the sector. But with great promise comes great responsibility. As this survey highlighted, the adoption of GenAI must be balanced with strong safeguards. Risks such as algorithmic bias, regulatory uncertainty, and emerging cyber threats cannot be overlooked. Encouragingly, a global wave of regulatory initiatives – from the EU AI Act to RBI's FREE-AI and MAS's MindForge – is laying the groundwork for ethical and accountable AI integration. Success in this AI-powered future will belong to institutions that treat GenAI not just as a tech upgrade but as a strategic transformation. By embedding governance, fostering interdisciplinary collaboration, and staying aligned with global best practices, they can turn GenAI into a true competitive edge. In short, the future of finance is not just AI-driven — it's *responsibly AI-driven*. Those who lead with both innovation and integrity will shape a more efficient, inclusive, and trusted financial ecosystem for the decade ahead.